\documentstyle[aps,preprint]{revtex}
\begin{document}
\title{Superconductivity from correlated hopping}
\author{C.D. Batista, F. Lema and A.A. Aligia}
\address{Comisi\'{o}n Nacional de Energ\'{i}a At\'{o}mica\\
Centro At\'{o}mico Bariloche and Instituto Balseiro\\
8400 Bariloche, Argentina}
\maketitle

\begin{abstract}
We consider a chain described by a next-nearest-neighbor hopping combined
with a nearest-neighbor spin flip. In two dimensions this three-body term
arises from a mapping of the three-band Hubbard model for CuO$_2$ planes to
a generalized $t-J$ model and for large O-O hopping favors
resonance-valence-bond superconductivity of predominantly $d$-wave symmetry.
Solving the ground state and low-energy excitations by analytical and
numerical methods we find that the chain is a Luther-Emery liquid with
correlation exponent $K_{\rho} = (2-n)^2/2$, where $n$ is the particle
density.
\end{abstract}

\medskip
PACS number: 74.20.-z, 71.27.+a, 75.10.Jm \newpage

Recently, there has been a considerable interest in obtaining the exact
ground state of different strongly correlated systems, using a variety of
techniques \cite{mon,geb,es1,es2,bra,st1,st2,arr1,ali1}. Although in some
cases the model or the parameters are rather unrealistic, exact results are
useful in clarifying the effect of different physical ingredients and as a
test of approximations. Particularly interesting are those models in which
the ground state is superconducting, or has dominant superconducting
correlations in one dimension \cite{es2,arr1}, because they display
superconducting mechanisms which are different from the conventional
electron-phonon one and might be relevant to the understanding of the high-$%
T_c$ systems.

Here we consider a one-dimensional (1D) system described by the Hamiltonian:
\begin{equation}
H_{t^{\prime\prime}} = t^{\prime\prime}\sum_{i \delta \neq
\delta^{\prime}\sigma} c^{\dagger}_{i + \delta^{\prime}\sigma} c_{i \delta
\sigma}(\frac{1}{2} - 2 {\bf S}_i \cdot {\bf S}_{i + \delta}),
\end{equation}
where in terms of ordinary fermion operators $c^{\dagger}_{i \sigma} = {%
\tilde c}^{\dagger}_{i \sigma} (1- \tilde c^{\dagger}_{i- \sigma} {\tilde c}%
_{i - \sigma})$ (double occupancy at any site is forbidden), ${\bf S}_i$ is
the spin operator at site $i$, and $i+ \delta$ denote the nearest neighbors
of site $i$. Eq.(1) can also be written in the form of a hopping of singlet
pairs:
\begin{equation}
H_{t^{\prime\prime}} = 2t^{\prime\prime}\sum_{i \delta \neq \delta^{\prime}}
b^{\dagger}_{i \delta^{\prime}} b_{i \delta} ; ~~ b^{\dagger}_{i \delta} =
\frac{1}{\sqrt{2}} (c^{\dagger}_{i + \delta \uparrow} c^{\dagger}_{i
\downarrow} - c^{\dagger}_{i + \delta \downarrow} c^{\dagger}_{i \uparrow})
\end{equation}

In 2D, $H_{t^{\prime\prime}}$ was obtained as an important term in a
systematic analytical derivation of a generalized $t-J$ model, starting from
the extended three-band Hubbard model \cite{ali2}. Also, a numerical study
of a Cu$_4$O$_8$ cluster has shown that fitting of the energy levels using a
generalized $t-J$ model improves considerably, and a correct ordering of the
levels is obtained, if $H_{t^{\prime\prime}}$ is included \cite{bati}. If
the direct O-O hopping $t_{pp}$ of the three-band model is large enough, $%
t^{\prime\prime}$ has opposite sign as the one which corresponds to a
canonical transformation of the Hubbard model \cite{ali2,bati2} and
simultaneously the direct nearest-neighbor hopping $t^{\prime}$ has the sign
which agrees with the observed Fermi surface and other properties of
hole-doped superconductors, and with the asymmetry in the magnetic
properties between hole-doped and electron-doped cuprates \cite
{ali2,bati2,chi,toh}.

A recent exact diagonalization of a 4${\times}$4 cluster shows that when $%
|t^{\prime\prime}/t| > 0.12$ with the sign of $t^{\prime\prime}$
corresponding to large $t_{pp}$, the system shows strong signals of $d$-wave
superconductivity for realistic $J/t = 0.4$ and doping $x = 1-n$ = 0.25 \cite
{bati2}. The $d$-wave superconducting susceptibility is strongly enhanced
and the ground-state wave function has a significant overlap with a a simple
short-range resonance-valence-bond state with superconducting off-diagonal
long-range order (ODLRO) in the $d$-wave and $s$-wave channels (the former
being the dominant). The fact that $H_{t^{\prime\prime}}$ changes the
physics dramatically is also evident in the symmetry of the ground state as
a function of doping \cite{bati2}. Another case of the three-band Hubbard
model in wich $H_{t^{\prime\prime}}$ is important is the limit $%
t_{pp}<<t_{pd}<<{\Delta}$ where ${\Delta}$ is the Cu-O charge transfer
energy. In this case $t^{\prime\prime}$ becomes of the order of $t$, as can
be seen from a canonical transformation that eliminates double occupancy in
the effective one-band Hubbard model \cite{sim}.

Unfortunately, although the saturation of the superconducting correlation
functions in Ref. \cite{bati2} is an indication of ODLRO, a study of a small
cluster cannot assure the presence of ODLRO in the thermodynamic limit.
Instead, in 1D one can use powerful results obtained using conformal-field
theory to relate the correlation exponents with thermodynamic quantities
\cite{con}, and then, the effect of $H_{t^{\prime\prime}}$ on
superconducting correlations can be determined unambiguosly. As a first
step, we drop the other terms of the generalized $t-J$ model. This allows us
to obtain analytical results and to isolate the effect of $%
H_{t^{\prime\prime}}$.

To gain physical insight into the 1D problem, let us consider a chain with
only one vacant site (or hole). The action of the Hamiltonian Eq.(2) over a
state in which the hole is localized at a given site, can be thought as a
permutation of the hole with a nearest-neighbor singlet. Thus, to take
maximum advantage of $H_{t^{\prime\prime}}$, it is convenient that the spin
configuration consists of a bond-ordering wave of non-overlaping singlets at
each side of the hole (see Fig. 1). The subsequent motion of the hole does
not alter the spin configuration, the problem becomes equivalent to a
tight-binding model with next-nearest-neighbor hopping $2t^{\prime\prime}$,
and the resulting ground-state energy in the thermodynamic limit $%
-4t^{\prime\prime}$ coincides with the lower bound of $H_{t^{\prime\prime}}$
which can be obtained using Gerschgorin's theorem \cite{ger,ali1}. This fact
confirms that the spin configuration is the optimum one. When more than one
hole is present in the system, in order to have all spins paired in singlets
of nearest-neighbor particles, it is necessary that the distance between two
successive holes is odd (the parity of this distance is conserved by $%
H_{t^{\prime\prime}}$). From the arguments given above, it seems clear that
the ground state is a combination of sequences of these singlets and holes
(see Fig.1). Although we could not prove analytically this statement for
more than two holes, we have verified that the ground-state energy of an
open chain with 14 sites and any even number of particles is the same as
that of a spinless model (Eqs.(3) and (4)), obtained using the mapping
sketched in Fig.1: each singlet is replaced by a spinless fermion. For a
chain of length $L$ and $N$ particles ($N$ even), the hopping, length and
number of particles of the effective spinless fermion problem are given by:
\begin{equation}
t = 2t^{\prime\prime}; L^{\prime}= L -N/2 ; N^{\prime}= N/2 .
\end{equation}
For example in Fig.1 $L=14$, $N=10$, and then $L^{\prime}=9$, $N^{\prime}=5$%
. In a chain with open boundary conditions, the order of the singlets is
conserved, and thus, the statistics of the fermions in the equivalent
problem does not play any role. We recall that the ground-state energy of a
spinless chain with open boundary conditions of length $L^{\prime}$, $%
N^{\prime}$ fermions and hopping $t$ is (see for example Ref. \cite{arr1}):
\begin{equation}
E = - |t| \left (\frac{{\rm sin} \pi \frac{2N^{\prime}+1}{2L^{\prime}+2}} {%
{\rm sin} \frac{\pi}{2L^{\prime}+2}} - 1 \right ).
\end{equation}

Eqs.(3) and (4) describe the ground-state energy of $H_{t^{\prime\prime}}$
for a chain with open boundary conditions. However, for the discussion below
concerning the central charge and low-energy excitations, it is more
convenient to know the energy of $H_{t^{\prime\prime}}$ in a ring (periodic
boundary conditions). In this case, the mapping to the fermionic system can
still be done: the fermionic sign which arises under a cyclic permutation of
the fermions can be absorbed taking periodic (antiperiodic) boundary
conditions if the number of fermions is odd (even). This corresponds to an
even number of particles not multiple of four (multiple of four) in the
original system described by $H_{t^{\prime\prime}}$. In both cases the
energy can be written as:
\begin{equation}
E = - 2 |t| \frac{{\rm sin} (\pi N^{\prime}/L^{\prime})}{{\rm sin} (\pi
/L^{\prime})}
\end{equation}
Using Eq.(3), the ground-state energy per site $e = E/L$ of a ring of length
$L$ and density $n = N/L$ described by $H_{t^{\prime\prime}}$ is:
\begin{equation}
e(n,L) = - \frac{4|t^{\prime\prime}|}{L} \frac{{\rm sin} ( \frac{\pi n}{2 - n%
})}{{\rm sin} (\frac{2 \pi}{L(2-n)})}
\end{equation}

We have also verified Eq.(6) for $L = 14$ and all $n$ multiple of 1/7. From
Eqs. (3) and (4) or Eq. (6) one obtains in the thermodynamic limit:
\begin{equation}
e(n, \infty) = - \frac{2|t^{\prime\prime}|}{\pi} (2-n) \sin (\frac{\pi n}{2-n%
})
\end{equation}
The corresponding wave function can be obtained through the mapping
procedure in a similar way as in Ref. \cite{arr1}.

To obtain the long-distance behavior of any correlation function, using
results of conformal-field theory, one needs information about the
low-energy excitations of the system \cite{con}. It is easy to see that the
low-energy triplet excitations on the ground state, act as barriers to the
propagation of the holes. From arguments similar to those presented above
for the one-hole case, one sees that the spin excitation of lowest energy is
to change one singlet of nearest-neighbor particles into a triplet. For open
boundary conditions in the equivalent fermionic problem, this is equivalent
to changing a chain of length $2L^{\prime}+1$ with $2N^{\prime}+1$ particles
into two equal chains of length $L$' whith $N^{\prime}$ particles. Using
Eqs. (3) and (4) we obtain that the value of the spin gap in the
thermodynamic limit is:
\begin{equation}
{\Delta}_s = |2t^{\prime\prime}| [1 + \cos ( \frac{\pi n}{2-n})]
\end{equation}
The gap is maximum for $n=0$ and vanishes at $n=1$.

The charge excitations, and in particular the charge velocity $v_c$ can also
be calculated using the equivalent model of spinless fermions. It should be
taken into account that distances and therefore also momentum are changed in
the mapping procedure. While the minimum excitation energy $E_{{\rm min}}$
of $H_{t^{\prime\prime}}$ in a ring corresponds to $q = 2 \pi/ L$, the
corresponding momentum for the same energy in the equivalent spinless ring
is $q^{\prime}= 2 \pi/L^{\prime}$. Calling $v_c^{\prime}$ the charge
velocity of the spinless ring, we can write \cite{arra2}:
\begin{equation}
v_c = \frac{E_{{\rm min}} - E}{2 \pi / L} = \frac{L}{L^{\prime}} \frac{E_{%
{\rm min}} - E}{2 \pi /L^{\prime}} = \frac{L}{L^{\prime}} v^{\prime}_c
\end{equation}
In the thermodynamic limit $v^{\prime}_c = 2t \sin (\pi
N^{\prime}/L^{\prime})$, and using Eqs.(3) and (9) one obtains:
\begin{equation}
v_c = \frac{8 |t^{\prime\prime}|}{2-n} \sin (\frac{\pi n}{2-n})
\end{equation}
Expanding the denominator of Eq.(6) up to third order in the argument and
using Eqs.(7) and (10), one can see that for $L \rightarrow \infty$:
\begin{equation}
e(n,L) = e(n, \infty) - c \frac{\pi v_c}{6L^2}
\end{equation}
where the number $c=1$ is called the central charge \cite{con,blo}. This
scaling law characterizes a Luther-Emery liquid \cite{lut}. In this liquid,
the correlation functions are characterized by a single exponent $K_{\rho}$.
The superconducting pair-pair correlation functions depend on the distance $d
$ between pairs as $d^{-1/K_{\rho}}$ for $d >> 1$ and are the dominant
correlation functions at large distances if $K_{\rho} > 1$ \cite
{con,lut,arra2}. In this case a small coupling between different chains
gives rise to superconductivity. $K_{\rho}$ can be calculated in terms of
the charge velocity and the compressibility $\kappa$ \cite{con,arra2}. Using
Eq.(7) on obtains:
\begin{equation}
\frac{1}{{\kappa}n^2} =\frac{\pi v_c}{2{K}_{\rho}} = \frac{\partial^2 e}{%
\partial n^2} =\frac{8 \pi|t^{\prime\prime}|}{(2-n)^3} \sin (\frac{\pi n}{2-n%
}),
\end{equation}
and using Eq.(10), a very simple expression for the correlation exponent
results:
\begin{equation}
{K }_{\rho} = \frac{(2-n)^2}{2}
\end{equation}

This exact result allows to test numerical methods and approximations. As an
example, we have calculated ${K}_{\rho}$ in a ring of 14 sites from the
expression ${K}_{\rho} = \pi({\kappa}n^2D_c/2)^{1/2}$, obtaining ${\kappa}$
and the Drude weight $D_c$ numerically as in Ref. \cite{arra2}. The
comparison with Eq.(13) is shown in Fig.2. One can see that the agreement is
very good except for a small deviation at the extreme densities. This
deviation is mainly due to the discretization of the expression of the
second derivative which enters the compressibility.

The system has dominant superconducting correlations for $0 < n < 2 -\sqrt{2}
$. Inspection of Eq.(12) shows that there is no phase separation in the
system for any density. This is in contrast to the $t-J$ and other models
proposed to the cuprates in which superconductivity takes place near a
region of phase separation \cite{ex}. In the dilute limit $n \rightarrow 0$,
$K_{\rho} \rightarrow 2$, as in the negative $U$ Hubbard model for large $|U|
$: the physics is like that of hard-core bosons which undergo Bose
condensation. The same type of physics is expected in 2D in the dilute
limit. We have confirmed this by numerical calculations in a 4$\times$4
cluster and comparison with a BCS-like wave function. Instead, for $n
\rightarrow 1$, $K_{\rho} \rightarrow 1/2$, as in the case of spinless
fermions. Addition of an attractive term like the superexchange $J$ of the $%
t-J$ model should extend the range of densities for which $K_{\rho} > 1$ to
larger values of $n$. One also expects that addition of $J$ should close the
spin gap for $n \sim 1$. The effect of the interplay of $H_{t^{\prime\prime}}
$ with the terms present in the $t-J$ model in 1D has been studied quite
recently by numerical methods \cite{lem,amm}. The main result is that the
regions of superconductivity and phase separation for $n \sim 1$ shift to
higher (lower) values of $J/t$ if a small $t^{\prime\prime}$ is added with
the same (opposite) sign as the one which corresponds to a canonical
transformation of the Hubbard model. Thus, as in 2D, the sign of $%
t^{\prime\prime}$ favored for large $t_{pp}$ in the mapping from the
three-band Hubbard model \cite{ali2,bati2} is more convenient for
superconductivity for realistic doping.

Two of us (C.D.B. and F.L.) are supported by the Consejo Nacional de
Investigaciones Cient\'{\i}ficas y T\'ecnicas (CONICET), Argentina. A.A.A.
is partially supported by CONICET. 

\newpage

\section*{Figure Captions}

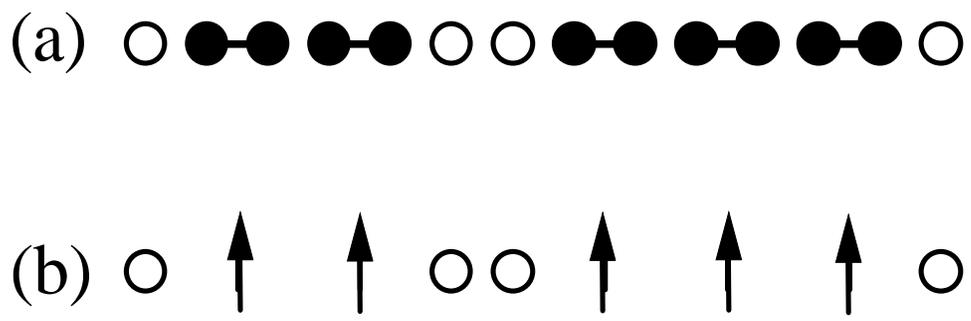
\noindent{\bf Fig. 1:}(a) Schematic representation of a typical state with
definite occupation at each site among those which compose the ground state
of the system. (b) state obtained from (a) replacing each singlet pair by a
spinless fermion. Open circles denote vacant sites, closed circles joined by
a full line represent a singlet $b^{\dagger}_{i \delta}$ (Eq. (2)), and up
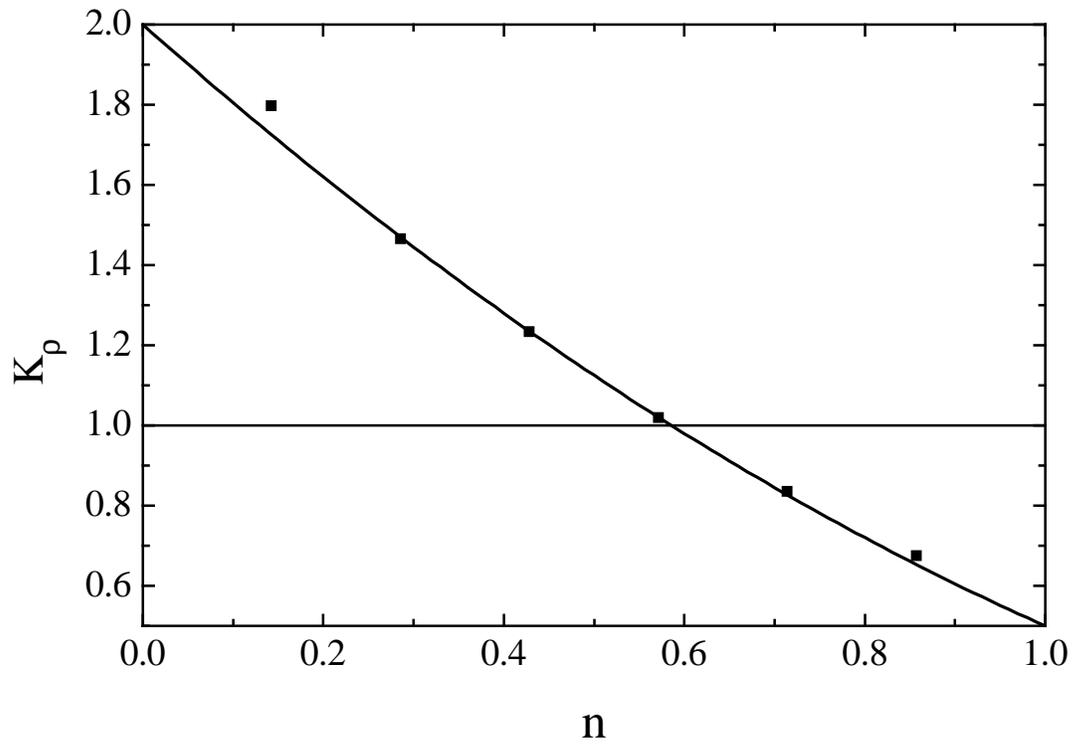
arrows represent spinless fermions. \\ \noindent{\bf Fig.2:} Correlation
exponent $K_{\rho}$ obtained numerically in a ring of 14 sites (solid
squares) in comparison with the exact result (solid line).

\end{document}